  \documentclass[preprint,showpacs,preprintnumbers,amsmath,amssymb]{revtex4}
\usepackage{graphicx}
\usepackage{dcolumn}
\usepackage{bm}
\usepackage{amssymb}
\usepackage{amsmath}

\begin{document}

\title{Peculiarities of sub-barrier fusion with quantum diffusion approach}
\author{V.V.Sargsyan$^1$, G.G.Adamian$^{1,2}$, N.V.Antonenko$^1$, and W. Scheid$^3$
}
\affiliation{$^{1}$Joint Institute for Nuclear Research, 141980 Dubna, Russia\\
$^{2}$Institute of Nuclear Physics, 702132 Tashkent, Uzbekistan\\
$^{3}$Institut f\"ur Theoretische Physik der
Justus--Liebig--Universit\"at,
D--35392 Giessen, Germany
}
\date{\today}

\begin{abstract}
With the quantum diffusion approach the unexpected behavior of fusion cross section,
angular momentum, and astrophysical $S$-factor at
sub-barrier energies has been revealed. Out of the region of short-range nuclear
interaction and action of friction at turning point the decrease rate of the cross section under the barrier becomes smaller.
The calculated results for the reactions with spherical nuclei
are in a good agreement with the existing experimental data.
\end{abstract}


\maketitle

\section{Introduction}
The fusion cross section at low energies crucially depends on the capture probability of the projectile by the
target nucleus, i.e. on the probability to pass the Coulomb barrier. There are many experimental and
theoretical studies of heavy ion fusion reactions at extreme sub-barrier energies.
The data obtained are of interest for solving the astrophysical problems related to nuclear synthesis.
Resent measurements of the fusion cross sections $\sigma$
at energies $E_{\rm c.m.}$ below the Coulomb barrier \cite{Ji1,Ji2,Dg,Mor}
showed the very rapid fall of $\sigma$ just below the barrier.
In the $S$-factor representation \cite{Zvezda}, $S=E_{\rm c.m.}\sigma \exp(2 \pi\eta)$ where
$\eta(E_{\rm c.m.})=Z_1Z_2e^2\sqrt{\mu/(2\hbar^2E_{\rm c.m.})}$
is the Sommerfeld parameter, the steep fall-off of the cross sections is related to a maximum of the
$S$-factor. The indications for this maximum have been found in Refs.~\cite{Ji2}.
However, its origin is still discussed. The so-called logarithmic
derivative, $L(E_{\rm c.m.})=d(\ln (\sigma E_{\rm c.m.}))/dE_{\rm c.m.}$, shows a growth at
$E_{\rm c.m.}$ corresponding to the maximum of $S$-factor.

The experiments \cite{Tr1,Og,Tr2} with the reactions $^{16}$O,$^{22}$Ne+$^{208}$Pb,
where the fusion and capture cross sections coincide, demonstrated the decreasing
rate of fall of the cross sections at energies about 3-4 MeV below the Coulomb
barrier. Although this finding has to be checked in other experiments, it can not
be presently ignored and deserves the theoretical analysis.
In this paper we will show that unexpected behavior of sub-barrier fusion
can be related to the switching off the nuclear interaction at external turning point $r_{ex}$.
If the colliding nuclei approach the distance $R_{int}$ between their centers, the
nuclear forces start to act in addition to the Coulomb interaction. Thus, at $R<R_{int}$
the relative motion may be more coupled with other degrees of freedom. At $R>R_{int}$
the relative motion is almost independent of the internal degrees of freedom.
Depending on whether the value of $r_{ex}$ is larger or smaller than interaction radius $R_{int}$,
the impact of coupling with other degrees of freedom upon the barrier passage
seems to be different.

To clarify the behavior of capture cross sections at sub-barrier
energies, the further development of the theoretical methods is required. The conventional coupled-channel
approach with realistic set of parameters is not able to describe the capture
cross sections either below or above the Coulomb barrier \cite{Dg}. The use of quite
shallow nucleus-nucleus potential \cite{Es} with adjusted repulsive core considerably
improves the agreement between the calculated and experimental data.
Besides the coupling with collective excitations,
the dissipation, which is simulated by an imaginary potential in Ref.~\cite{Es} or
by damping in each channel in Ref.~\cite{Hag}, seems to be important.

The quantum diffusion approach \cite{our} is based on the quantum master-equation for
the reduced density matrix and takes into account the fluctuation and dissipation effects in
collisions of heavy ions which model the coupling with various channels.
As demonstrated in Ref.~\cite{our}, this approach is
successful for describing the capture cross sections at energies near the
Coulomb barrier. Here, we apply it at wide energy interval including the
extreme sub-barrier region. To avoid the effects of nuclear deformation,
we treat  reactions  $^{16}$O,$^{22}$Ne,$^{48}$Ca+$^{208}$Pb.
The collisions of these spherical nuclei is treated in terms of a
single collective variable: the relative distance $R$ between the
colliding nuclei and the conjugate momentum $P$. The coupling with
other degrees of freedom is taken into account in the friction and diffusion.

\section{Model}
We use the nucleus-nucleus
potential \cite{poten} which naturally contains the repulsive part because of a
density-dependent nucleon-nucleon force. The nucleon densities of the projectile and target nuclei
are specified in the form of the Woods-Saxon parameterization,
where the nuclear radius parameter is $r_0=1.15$  fm  and the
diffuseness parameter takes the values $a= 0.55$ fm  for the $^{208}$Pb, $^{48}$Ca, $^{22}$Ne
nuclei and $a= 0.53$ fm  for the $^{16}$O nucleus.
The nucleus-nucleus potential $V$ has quite a shallow pocket (Fig.~1).
As the centrifugal part of the potential grows, the pocket
depth becomes smaller, while the position of the pocket minimum moves towards the barrier at $R=R_b$.
This pocket is washed out at angular momenta $J>80$.
As demonstrated in Refs.\cite{Es,Hag1}, the internuclear potentials
show a significant deviation from the conventional Woods-Saxon shape.
 The thicker the
potential is, the smaller the penetrability is, and also the
stronger the energy dependence of the penetrability is. The
thick potential barrier obtained for the $^{16}$O+$^{208}$Pb
reaction is thus consistent with the recent experimental
observations \cite{Es} that the fusion excitation function is much
steeper than theoretical predictions at deep sub-barrier energies.
The present study suggests
that the origin of the steep fall-off of fusion cross
section is partly attributed to the deviation of
the internuclear potential from the Woods-Saxon shape.

The capture cross section is a sum of partial capture cross sections
\begin{eqnarray}
\sigma_{c}(E_{\rm c.m.})&=&\sum_{J}^{}\sigma_{\rm c}(E_{\rm
c.m.},J)\nonumber\\
&=& \pi\lambdabar^2 \sum_{J}^{}(2J+1)P_{\rm cap}(E_{\rm
c.m.},J),
\label{1a_eq}
\end{eqnarray}
where $\lambdabar^2=\hbar^2/(2\mu E_{\rm c.m.})$ is the reduced de Broglie wavelength
and the summation is in possible values of angular momentum $J$
at given bombarding energy $E_{\rm c.m.}$. The partial capture probability
$P_{\rm cap}$ is defined by the passing probability of the potential barrier
in the relative distance $R$ at given $J$.

The value of $P_{\rm cap}$
can be obtained by integrating the propagator $G$ from the initial
state $(R_0,P_0)$ at time $t=0$ to the final state $(R,P)$ at time $t$:
\begin{eqnarray}
P_{\rm cap}&=&\lim_{t\to\infty}\int_{-\infty}^{r_{\rm in}}dR\int_{-\infty}^{\infty}dP\  G(R,P,t|R_0,P_0,0)\nonumber \\
&=&\lim_{t\to\infty}\frac{1}{2} {\rm erfc}\left[\frac{-r_{\rm in}+\overline{R(t)}}
{{\sqrt{\Sigma_{RR}(t)}}}\right].
\label{1ab_eq}
\end{eqnarray}
The second line in (\ref{1ab_eq}) is obtained by using the propagator
$G=\pi^{-1}|\det {\bf \Sigma}^{-1}|^{1/2}
\exp(-{\bf q}^{T}{\bf \Sigma}^{-1}{\bm q})$
($q_R(t)=R-\overline{R(t)}$, $q_P(t)=P-\overline{P(t)}$, $\overline{R(t=0)}=R_0$,
$\overline{P(t=0)}=P_0$, $\Sigma_{ij}(t)=2\overline{q_i(t)q_j(t)}$, $\Sigma_{ij}(t=0)=0$,
$i,j=R,P$) calculated in Ref.~\cite{DMDadonov} for
the inverted oscillator which approximates
the nucleus-nucleus potential $V$ in the variable $R$.
The frequency $\omega$ of
this oscillator with internal turning point $r_{\rm in}$ is defined from the condition of equality of the
classical actions of approximated and realistic potential barriers of the same hight at given $J$.
It should be noted that the passage through the Coulomb barrier approximated by a parabola
has been previously studied in Refs.~\cite{Hofman,VAZ,Ayik}. This
approximation seems to be well justified for the reactions considered.

All quantum-mechanical, dissipative  effects and non-Markovian effects accompanying
the passage through the potential barrier are taken into consideration in our formalism \cite{our}.
 The generalized fluctuation-dissipation
relations contain the influence of quantum effects on the collective motion.
We address
the dynamics of damped non-Markovian collective mode in terms of the first moment $\overline{R(t)}$
(the average of the coordinate $R$)
and variance $\Sigma_{RR}(t)$  in the coordinate:
\begin{eqnarray}
\overline{R(t)}&=&A_tR_0+B_tP_0,\nonumber\\
\Sigma_{RR}(t)&=&\frac{2\hbar^2\tilde\lambda\gamma^2}{\pi}
\int\limits_{0}^{t}
d\tau^{'}B_{\tau^{'}} \int\limits_{0}^{t} d\tau^{''}B_{\tau^{''}}
\int\limits_{0}^{\infty} d\Omega \frac{\Omega}{\Omega^2+\gamma^2}\nonumber\\
&\times&\coth\left[\frac{\hbar\Omega}{2T}\right]
\cos[\Omega (\tau^{'}-\tau^{''})],\nonumber\\
B_t&=&\frac{1}{\mu}\sum_{i=1}^{3}\beta_i(s_i+\gamma)e^{s_it},\nonumber\\
\quad A_t&=&\sum_{i=1}^{3}\beta_i[s_i(s_i+\gamma)+\hbar\tilde\lambda\gamma/\mu]e^{s_it}.
\label{RZ_eq}
\end{eqnarray}
The derivation of equations for $\overline{R(t)}$ and
 $\Sigma_{RR}(t)$ is presented in Refs.\cite{our,VAZ}
 as well as in Appendix A. Here, $\overline{R(0)}=R_0$, $\Sigma_{RR}(0)=0$,
 $A_0=1$, and $B_0=0$.
In Eqs.(\ref{RZ_eq}),
$\beta_1=[(s_1-s_2)(s_1-s_3)]^{-1}$,  $\beta_2=[(s_2-s_1)(s_2-s_3)]^{-1}$ and
$\beta_3=[(s_3-s_1)(s_3-s_2)]^{-1}$, and
$s_i$ are the real roots ($s_1\ge 0> s_2 \ge s_3$) of the following equation:
\begin{eqnarray}
 (s+\gamma)(s^2- \omega_0^2)+\hbar\tilde\lambda\gamma s/\mu=0.
 \label{Root_eq}
\end{eqnarray}
Here, $\mu$ is the reduced mass, $\omega_0^2=\omega^2\{1-\hbar\tilde\lambda\gamma/
[\mu(s_1+\gamma)(s_2+\gamma)]\}$ is  the renormalized frequency in the Markovian limit,
and the value of
$\tilde\lambda$ is related to the strength of linear coupling
in coordinates between collective and internal subsystems.
The friction coefficient in $R$ is set as $\hbar\lambda=-\hbar (s_1+s_2)=2$ MeV. It
has the value close to those calculated within other approaches \cite{Den}.
Because of the quite large values of $\omega$, the calculated results
are rather insensitive to the value of temperature $T<1$ MeV.
Non-Markovian effects appear in the calculations
through the internal-excitation width $\hbar\gamma= 15$ MeV.
Since the relaxation time for the internal subsystem
is much shorter than the characteristic time of collective motion, the condition
$\gamma\gg \omega$ should be fulfilled. The value of $\tilde\lambda$ is partly related
to the value of $\gamma$. In the limit of $\tilde\lambda\to 0$
the value of $\gamma$ should go to infinity to meet the Markovian dynamics.

The Eqs.~(\ref{1ab_eq}) and (\ref{RZ_eq}) result
\begin{eqnarray}
P_{\rm cap}&=&
\frac{1}{2} {\rm erfc}\left[\left(\frac{s_1(\gamma-s_1)}{2\hbar\tilde\lambda \gamma}\right)^{1/2}\right.\nonumber\\
&\times& \left.\frac{\mu\omega_0^2 R_0/s_1+P_0}
{\left[\frac{s_1\gamma}{\pi(s_1+\gamma)}
\left(\psi(1+\frac{\gamma}{2\pi T})-\psi(\frac{s_1}{2\pi T})\right)-T\right]^{1/2}}\right],
\end{eqnarray}
where $\psi(z)$ is digamma function.
Since  $\Sigma_{RR}\to \infty$ at $t\to \infty$, the value
of $P_{\rm cap}$ is independent of $r_{\rm in}$. Using Eq.~(\ref{Root_eq}),
in the limit of small temperature ($T\to 0$),
which is suitable for sub-barrier fusion, we obtain
\begin{eqnarray}
P_{\rm cap}&=&
\frac{1}{2} {\rm erfc}\left[\left(\frac{\pi s_1(\gamma-s_1)}{2\mu(\omega_0^2-s_1^2)}\right)^{1/2}
\frac{\mu\omega_0^2 R_0/s_1+P_0}
{\left[\gamma \ln(\gamma/s_1)\right]^{1/2}}\right].
\label{PC_eq}
\end{eqnarray}
This expression is used to calculate $P_{\rm cap}$ in our paper.
The conventional approach to modeling fusion cross sections is based on calculating quantum mechanical
barrier transmission probabilities for each partial wave of relative motion,
without regard (in first approximation) to coupling to internal degrees of freedom.
If the coupling with internal degrees of freedom is disregarded in the quantum diffusion approach at
zero temperature, then
$\tilde\lambda \to 0$, $s_1\to \omega_0$, and $\gamma \to \infty$
so that $4\hbar \tilde\lambda \ln(\gamma/\omega_0)\to 1$.
In this case the well-known expression $P_{\rm cap}\sim \exp[-2 \pi (V_b-E_{\rm c.m.})/(\hbar\omega_0)]$
is obtained where $V_b$ is the hight of the barrier at given $J$. Thus, our diffusion approach contains
the quantum mechanical barrier transmission probability.
In the presence of coupling with internal degrees of freedom
the capture at sub-barrier energies also occurs due to the quantum noise.
The actions of friction and diffusion are the reverse of each other \cite{our}
in the dissipation of energy. As a result, in the reactions treated
the dissipation of energy is negligible at $V_b-E_{\rm c.m.}>3.5$ MeV.

The nuclear forces start to play a role
at $R_{int}=R_b+1.1$ fm where the nucleon density of colliding nuclei approximately reaches
10\% of saturation density. In Fig.~2 the interaction radius $R_{int}$ as well
as internal and external turning points are shown for the $^{16}$O+$^{208}$Pb
reaction at zero angular momentum. If the value of $r_{\rm ex}$ corresponding to external turning point
is larger than interaction radius $R_{int}$,
we take $R_0=r_{\rm ex}$ and $P_0=0$ in Eq.~(\ref{PC_eq}).
For $r_{\rm ex}< R_{int}$, it is naturally
to start our treatment with $R_0=R_{int}$ and $P_0$ defined by the kinetic energy
at $R=R_0$. In this case the friction hinders the classical motion towards smaller $R$.
If $P_0=0$ at $R_0>R_{int}$, the friction almost does not play a role in the transition
through the barrier.
Thus, two regimes of interaction at sub-barrier energies differ by the action
of nuclear forces and the role of friction at $R=r_{\rm ex}$.

\section{Calculated results}
Besides the parameters related to the nucleus-nucleus potential,
two parameters $\hbar\gamma=15$ MeV and $\hbar\lambda=2$ MeV are
used for calculating the capture probability. All calculated results are
obtained with the same set of parameters and
are rather insensitive to the reasonable variation of them \cite{our,VAZ}.

In Figs.~3-5 the calculated capture cross sections for the reactions
$^{16}$O,$^{22}$Ne,$^{48}$Ca+$^{208}$Pb are in a good agreement
with the available experimental data \cite{Dg,Mor,Tr1,Og,Tr2,Bock,Ca1,Ca2}.
There is sharp fall-off of the cross sections
just under the barrier. With decreasing $E_{\rm c.m.}$ up to about 3.5-5.0 MeV below
the Coulomb barrier the regime of interaction is changed because at external
turning point the colliding nuclei do not reach the region of nuclear interaction
where the friction plays a role.
As a result, at smaller $E_{\rm c.m.}$ the cross sections fall with smaller rate.
With larger value of $R_{int}$ the change of fall rate would occur at smaller $E_{\rm c.m.}$.
However, the uncertainty in definition of $R_{int}$ is rather small. Therefore, the
effect of the change of fall rate of sub-barrier fusion cross section should be in the data
if we believe that friction start to act only when the colliding nuclei approach the barrier.
Note that at energies of 5 MeV below the barrier the experimental data
have still large uncertainties to make a firm experimental conclusion about
this effect.
The effect seems to be more pronounced in the collisions of spherical nuclei. The collisions
of deformed nuclei occurs at various mutual orientations from which the value of $R_{int}$
depends.

The calculated average angular momenta $\langle J^2\rangle =\sum_J J(J+1) \sigma_c(E_{\rm c.m.},J)/
\sigma_c(E_{\rm c.m.})$ of fused systems versus $E_{\rm c.m.}$ are presented in Figs.~3-5 as well.
At energies of 3-4.5 MeV below the barrier $\langle J^2\rangle$ has a minimum. The
experimental data \cite{Vand} indicate the presence of the minimum as well. On the left-hand side
of this minimum the dependence of $\langle J^2\rangle$ of $E_{\rm c.m.}$ is rather weak.
The similar weak dependence has been found in Ref.~\cite{Bala} at extreme sub-barrier region.
Note that the found behavior of $\langle J^2\rangle$, which is
related to the change of the regime of interaction between the colliding nuclei, would affect the angular
anisotropy of the products of fission following fusion.

In Fig.~6 the functions $L(E_{\rm c.m.})$ and $S(E_{\rm c.m.})$, and
the fusion barrier distribution $d^2(E_{\rm c.m.}\sigma)/d E_{\rm c.m.}^2$ are
presented for the $^{16}$O+$^{208}$Pb reaction. The logarithmic derivative strongly
increases just below the barrier and then has a maximum. This leads to the
maximum of $S$-factor which is seen in the experiments \cite{Es}. After this maximum
$S$-factor decreases with $E_{\rm c.m.}$ and then starts to increase.
The same behavior has been revealed in Refs. \cite{LANG} by extracting the $S$-factor from
the experimental data.
If for finding the logarithmic derivative we use
only the cross sections calculated at the energies where the experimental
cross sections are available, the function $L(E_{\rm c.m.})$ would
be similar to that obtained with the experimental data \cite{Dg}. Therefore,
the energy increment in Ref.~\cite{Dg} seems to be too large
to extract a function $L(E_{\rm c.m.})$ with a very narrow maximum. This increment should be
at least 0.2 MeV. The fusion barrier distribution calculated
with this small energy increment has only one maximum. Using
larger energy increment of 0.6 MeV, one can get few oscillations in
$d^2(E_{\rm c.m.}\sigma)/d E_{\rm c.m.}^2$.

\section{Summary}
The quantum diffusion approach has been applied to study the
capture or fusion cross sections at sub-barrier energies. The
available experimental data at energies above and below the Coulomb barrier
are well described. Due to the change of the regime of interaction
(the turning-off the nuclear forces and friction)
at sub-barrier energies, the decrease rate of the cross sections is changed
at about 3.5-5.0 MeV below the barrier.
This change is reflected in the functions $\langle J^2\rangle$, $L(E_{\rm c.m.})$, and $S(E_{\rm c.m.})$.
The average angular momentum of compound nucleus versus $E_{\rm c.m.}$ would have a minimum
and then saturation at sub-barrier energies.
This behavior of $\langle J^2\rangle$ would increase the expected anisotropy
of angular distribution of the products of fission following fusion.
The energy increment of 0.2 MeV has to be used
in the experiment to get the cross sections suitable for calculating the value of $L$ and
the barrier distribution.

This work was supported by DFG and RFBR. The Polish-JINR and IN2P3-JINR Cooperation
programs are gratefully acknowledged.

\appendix
\section{}
For quantum nuclear system, the Hamiltonian $H$ depending
explicitly on the collective coordinate $R$, the canonically conjugate collective momentum $P$,
and on the internal
degrees of freedom has been constructed in  Refs.~\cite{our,VAZ}.
Using this Hamiltonian, one can obtain the system of integro-differential stochastic equations
for the Heisenberg operators $R$ and $P$
\begin{eqnarray}
\dot R(t)&=&\frac{P(t)}{\mu},\nonumber\\
\dot P(t)&=&\mu\omega_0^2 R(t) -
\int\limits_{0}^{t}d\tau K(t-\tau)\dot R(\tau) +  F(t).
\label{4_eq}
\end{eqnarray}
Here,
 $K(t)=\hbar\tilde\lambda\gamma e^{-\gamma t}$
is a dissipation kernel and $F(t)$ is a random  force.
This system of Eqs. (A1) is  the system of generalized nonlinear Langevin equations.
The integral term in the equations
of motion means that the system is non-Markovian and has a "memory" of the motion in the trajectory
preceding the instant $t$.
The operator of the random force $F$ has the form of a Gaussian distribution with a zero mean,
$\langle\langle F(t) \rangle\rangle = 0$, and a nonzero variance,
\begin{eqnarray}
\langle\langle F(t)F(t') \rangle\rangle =\frac{\hbar^2\tilde\lambda\gamma^2}{\pi}
\int\limits_{0}^{\infty}
d\Omega \frac{\Omega}{\Omega^2+\gamma^2}\coth\left[\frac{\hbar\Omega}{2T}\right]\cos[\Omega (t-t^{'})].
\end{eqnarray}
The symbol $\langle\langle\, .\, \rangle\rangle$ denotes the mean over the  internal
degrees of freedom.
To solve Eqs. (A1) analytically, we  use the Laplace transform method.
After finding expressions for the
images, we obtain explicit expressions for the originals ~\cite{our,VAZ},
\begin{eqnarray}
R(t)=A_tR(0)+B_tP(0)+\int\limits_{0}^{t}d\tau
B_\tau F(t-\tau),
\end{eqnarray}
where the explicit expressions for $A_t$ and $B_t$ are given in Eq.(3).
Using the time dependence $R(t)$, we obtain the
values $\overline{R(t)}$ (see Eq.(3)) and
\begin{eqnarray}
\Sigma_{RR}(t)=2\overline{(R(t)-\overline{R(t)})^2}=
2\int\limits_{0}^{t}
d\tau^{'}B_{\tau^{'}} \int\limits_{0}^{t} d\tau^{''}B_{\tau^{''}}\langle\langle F(t-\tau^{'})F(t-\tau^{''})\rangle\rangle
\end{eqnarray}
averaged over the whole system. Using Eq.(A2), Eq.(A4) can be rewritten
in the form of Eq.(3).

\begin{figure}
\caption{The nucleus-nucleus potentials calculated at $J=0$ (solid curve), 30 (dashed curve),
60 (dotted curve), and 90 (dash-dotted curve) for the $^{16}$O+$^{208}$Pb reaction.
}
\label{1_fig}
\end{figure}

\begin{figure}
\caption{The nucleus-nucleus potential calculated at $J=0$ (solid lines) for the $^{16}$O+$^{208}$Pb reaction.
The position $R_b$ of the Coulomb barrier, radius of interaction $R_{int}$, and external
and internal turning points for some value of $E_{\rm c.m.}$ are indicated.
}
\label{2_fig}
\end{figure}

\begin{figure}
\caption{The calculated (solid lines) fusion cross section versus $E_{\rm c.m.}$ (upper part) and
average angular momenta of compound nucleus (lower part) versus $E_{\rm c.m.}$
for the $^{16}$O+$^{208}$Pb reaction are compared with the experimental data. The experimental
cross sections marked by open squares and circles, and closed rhombus and triangles are
from Refs.~\protect\cite{Dg,Mor,Tr1,Og}, respectively. The experimental
values of $\langle J^2\rangle$ (solid squares) are from Ref.~\protect\cite{Vand}.
The value of the Coulomb barrier $V_b$ is indicated by arrow.
}
\label{3_fig}
\end{figure}

\begin{figure}
\caption{The same as in Fig.~3, but for the $^{22}$Ne+$^{208}$Pb reaction.
The experimental data marked by closed squares and circles are
from the two runs of Ref.~\protect\cite{Tr2}.
}
\label{4_fig}
\end{figure}

\begin{figure}
\caption{The same as in Fig.~3, but for the $^{48}$Ca+$^{208}$Pb reaction.
The experimental data marked by closed circles and squares, and open circles are
taken from Refs.~\protect\cite{Bock,Ca1,Ca2}, respectively.
}
\label{5_fig}
\end{figure}

\begin{figure}
\caption{The calculated values (solid lines) of logarithmic derivative $L$ (upper part),
astrophysical $S$-factor (middle part) with $\eta_0=\eta(E_{\rm c.m.}=V_b)$,
and fusion barrier distribution (lower part)
for the $^{16}$O+$^{208}$Pb reaction. The dashed line shows the values of $L$ obtained only with
the calculated cross sections at $E_{\rm c.m.}$ used in the experiment \protect\cite{Dg}.
}
\label{6_fig}
\end{figure}


\begin{thebibliography}{99}
\bibitem{Ji1}     C.L.~Jiang {\it et al.},   Phys. Rev. Lett. {\bf 89}, 052701 (2002);
C.L.~Jiang {\it et al.},   Phys. Rev. Lett. {\bf 93}, 012701 (2004).
\bibitem{Ji2}     C.L.~Jiang {\it et al.},   Phys. Rev. C {\bf 71}, 044613 (2005);
                  C.L.~Jiang, H.~Esbensen, B.B.~Back, R.V.F.~Janssens, and
K.E.~Rehm,   Phys. Rev. C {\bf 69}, 014604 (2004).
\bibitem{Dg}     M.~Dasgupta {\it et al.}, 
                 Phys. Rev. Lett. {\bf 99}, 192701 (2007).
\bibitem{Mor}     C.R.~Morton {\it et al.},   Phys. Rev. C {\bf 60}, 044608 (1999).
\bibitem{Zvezda}  K.~Langanke and  C.A. Barnes,   Adv. Nucl. Phys.  {\bf 22}, 173 (1996).
\bibitem{Tr1}     S.P.~Tretyakova, A.A.~Ogloblin, R.N.~Sagaidak,
                S.V.~Khlebnikov, and W.~Trzaska,   Nucl. Phys. {\bf A734}, E33 (2004).
\bibitem{Og}     Yu.Ts.~Oganessian {\it et al.},  JINR Rapid Communications {\bf 75}, 123 (1996).
\bibitem{Tr2}     S.P.~Tretyakova, A.A.~Ogloblin, R.N.~Sagaidak, W.~Trzaska, S.V.~Khlebnikov, R.~Julin, and J.~Petrowski,
                   Nucl. Phys. {\bf A738}, 487 (2004).
\bibitem{Es}     H.~Esbensen and C.L.~Jiang,   Phys. Rev. C {\bf 79}, 064619 (2009);
                  S.~Misicu and H.~Esbensen,   Phys. Rev. C {\bf 75}, 034606 (2007);
                  H.~Esbensen and S.~Misicu, Phys. Rev. C {\bf 76}, 054609 (2007).

\bibitem{Hag}     T.~Ichikawa, K.~Hagino, and A.~Iwamoto,  Phys. Rev. Lett. {\bf 103}, 202701 (2009);
                   K.~Hagino, N.~Rowley, and A.T.~Kruppa, Comput. Phys. Commun. {\bf 123}, 143 (1999).

\bibitem{our}     V.V.~Sargsyan, Z.~Kanokov, G.G.~Adamian, N.V.~Antonenko, and W.~Scheid,
                  Phys. Rev. C {\bf 80}, 034606 (2009); Phys. Rev. C {\bf 80}, 047603 (2009).

\bibitem{poten}     G.G.~Adamian {\it et al.}, Int. J.  Mod. Phys. E {\bf 5}, 191 (1996).

\bibitem{Hag1}     K.~Hagino and N.~Rowley, AIP Conf. Proc. {\bf 1098}, 18 (2009).
\bibitem{DMDadonov} V.V. Dodonov and V.I. Man'ko, Trudy Fiz. Inst. AN {\bf 167}, 7 (1986).

\bibitem{Hofman}    H.~Hofmann, Phys. Rep.  {\bf 284}, 137 (1997);
                    C.~Rummel and H.~Hofmann, Nucl. Phys. A {\bf 727}, 24 (2003).
\bibitem{VAZ}       G.G. Adamian, N.V. Antonenko, Z. Kanokov, and V.V. Sargsyan, Teor. Mat. Fiz. {\bf 145}, 87 (2005)
                    [Theor. Math. Phys. {\bf 145}, 1443 (2006)];
                    Z. Kanokov, Yu.V. Palchikov,  G.G. Adamian, N.V. Antonenko, and W. Scheid, Phys. Rev. E {\bf 71}, 016121 (2005);
                    Yu.V. Palchikov, Z. Kanokov,  G.G. Adamian, N.V. Antonenko, and W. Scheid, Phys. Rev. E {\bf 71}, 016122 (2005).
\bibitem{Ayik}      N.~Takigawa, S.~Ayik, K.~Washiyama, and S.~Kimura, Phys. Rev. C {\bf 69}, 054605 (2004);
 S.~Ayik, B.~Yilmaz, A. Gokalp, O. Yilmaz, and N.~Takigawa, Phys. Rev. C {\bf 71}, 054611 (2005).

\bibitem{Den} K.~Washiyama, D.~Lacroix, and S.~Ayik, Phys. Rev. C {\bf 79}, 024609
(2009); S.~Ayik, K.~Washiyama, and D.~Lacroix, {\it ibid}. {\bf 79}, 054606 (2009).

\bibitem{Bock}     R.~Bock {\it et al.}, Nucl. Phys. {\bf A388}, 334 (1982).

\bibitem{Ca1} A.J.~Pacheco {\it et al.}, Phys. Rev. C {\bf 45}, 2861 (1992).

\bibitem{Ca2}     E.~Prokhorova {\it et al.}, Nucl. Phys. {\bf A802}, 45 (2008).

\bibitem{Vand}    R.~Vandenbosch, Annu. Rev. Nucl. Part. Sci. {\bf 42}, 447 (1992).

\bibitem{Bala}    A.B.~Balantekin, J.R.~Bennett, and S.~Kuyucak, Phys. Lett. B {\bf 335}, 295 (1994).
\bibitem{LANG}    K.~Langanke and  S.E. Koonin,   Nucl. Phys.  {\bf A410}, 334 (1983);
A.Redder  {\it et al.}, Nucl. Phys. {\bf A462}, 385 (1987).
\end{thebibliography}
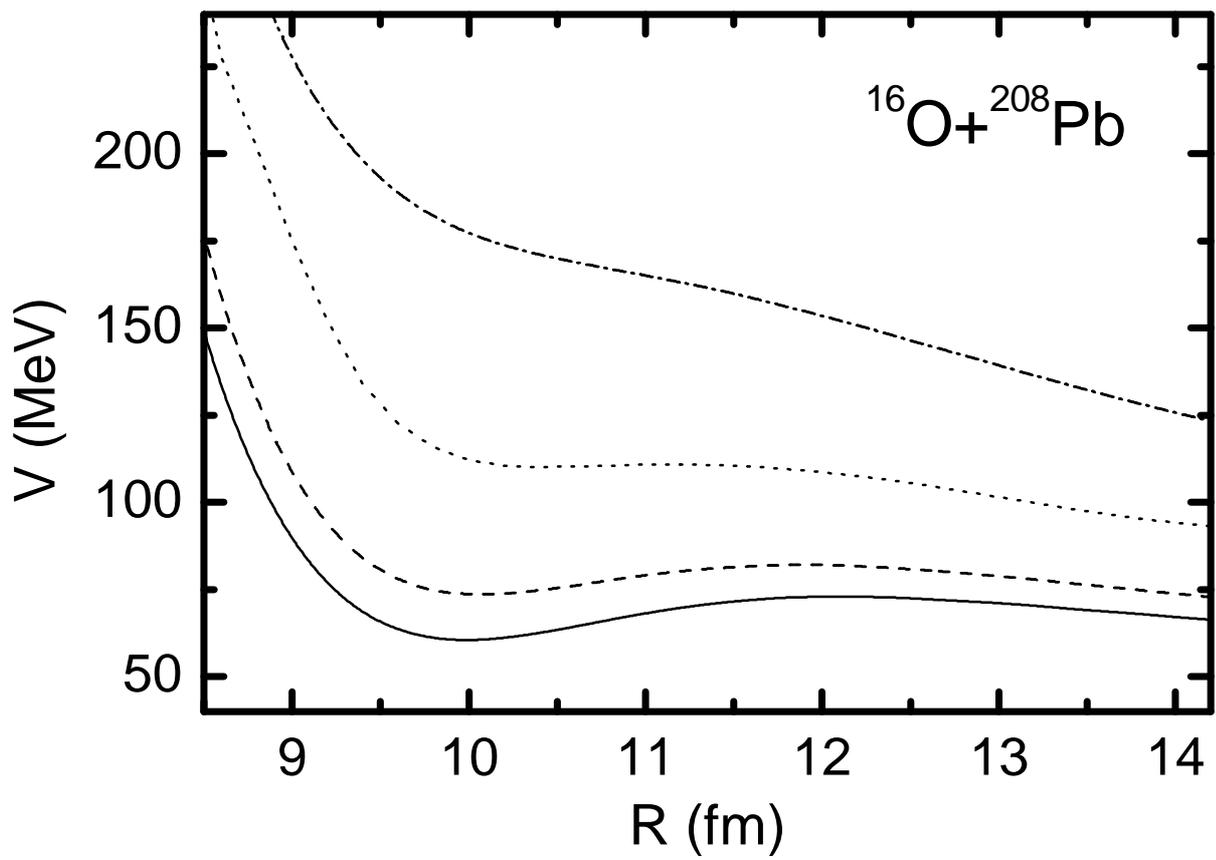
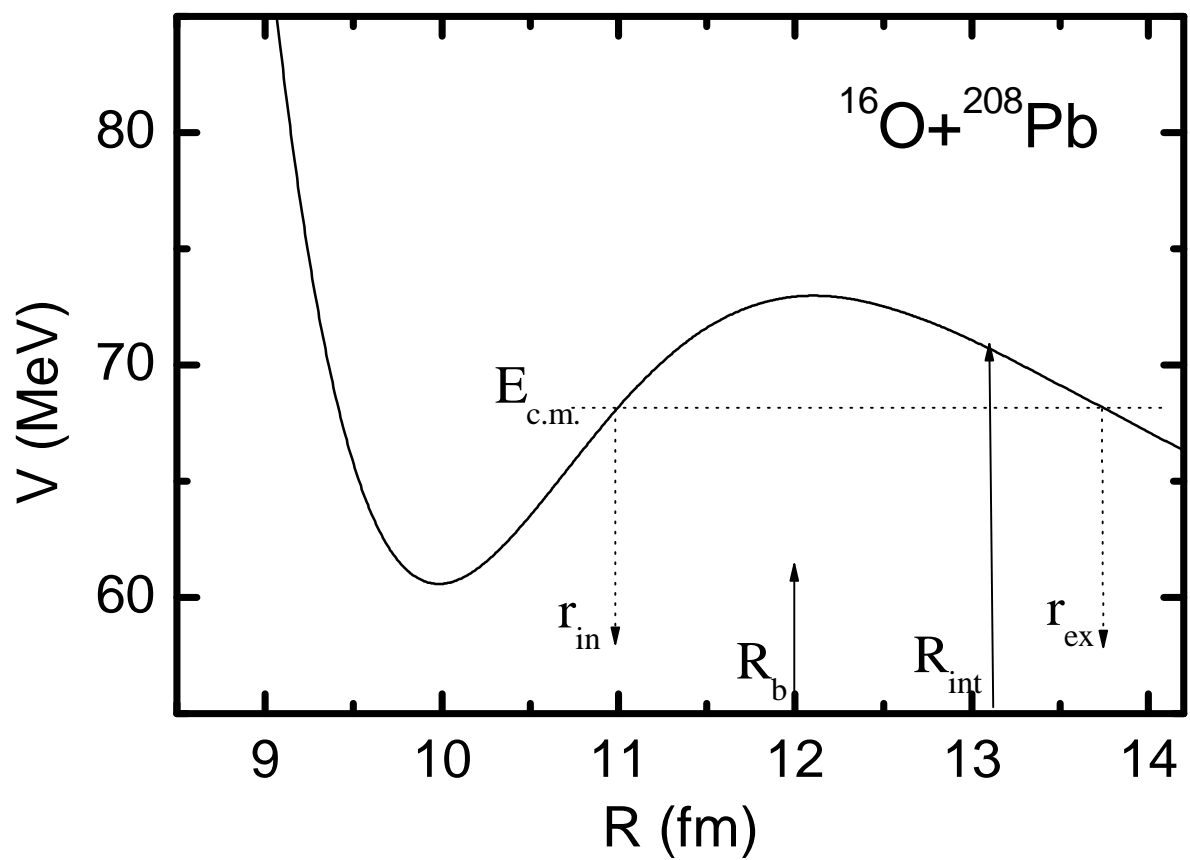
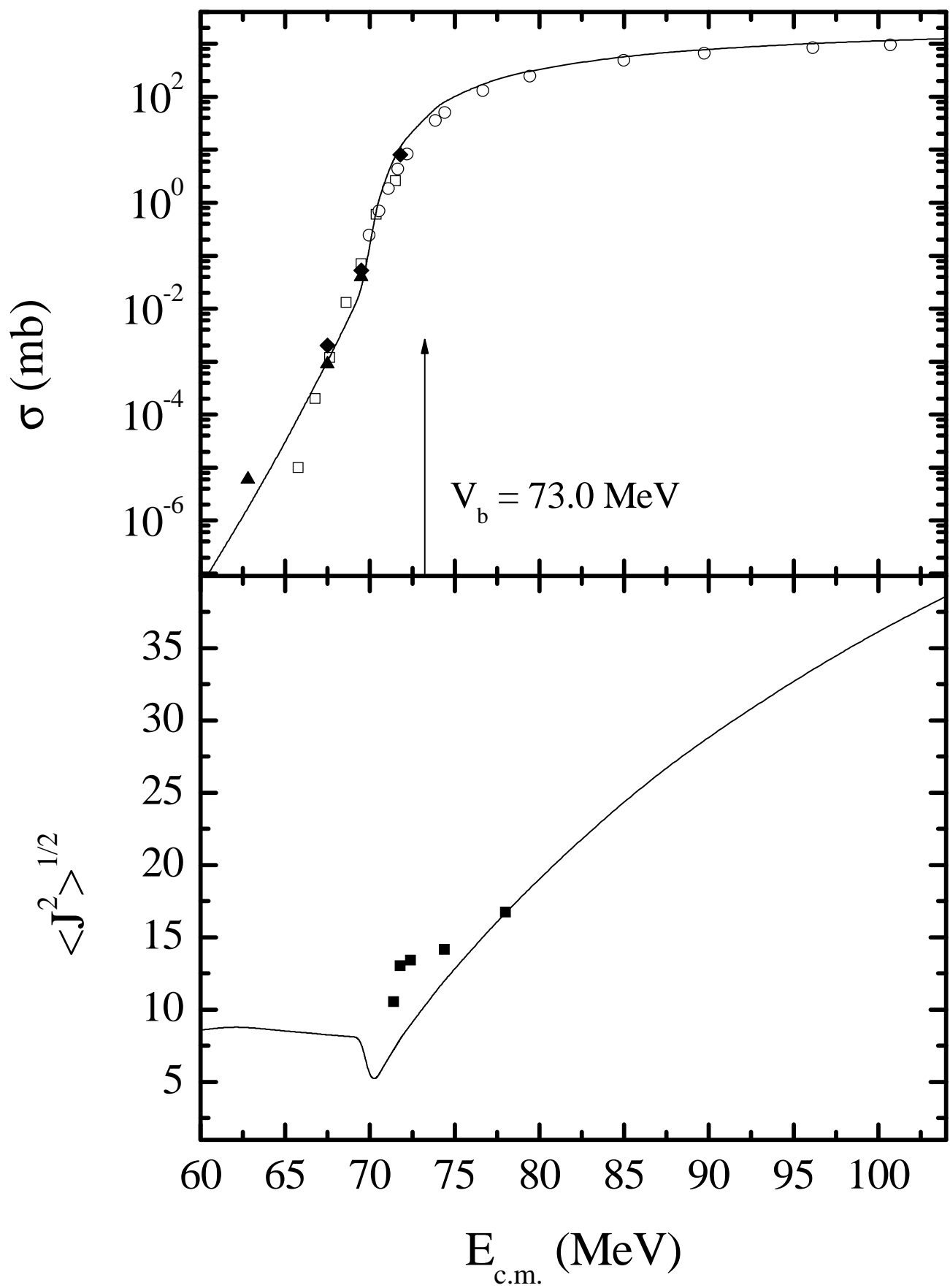
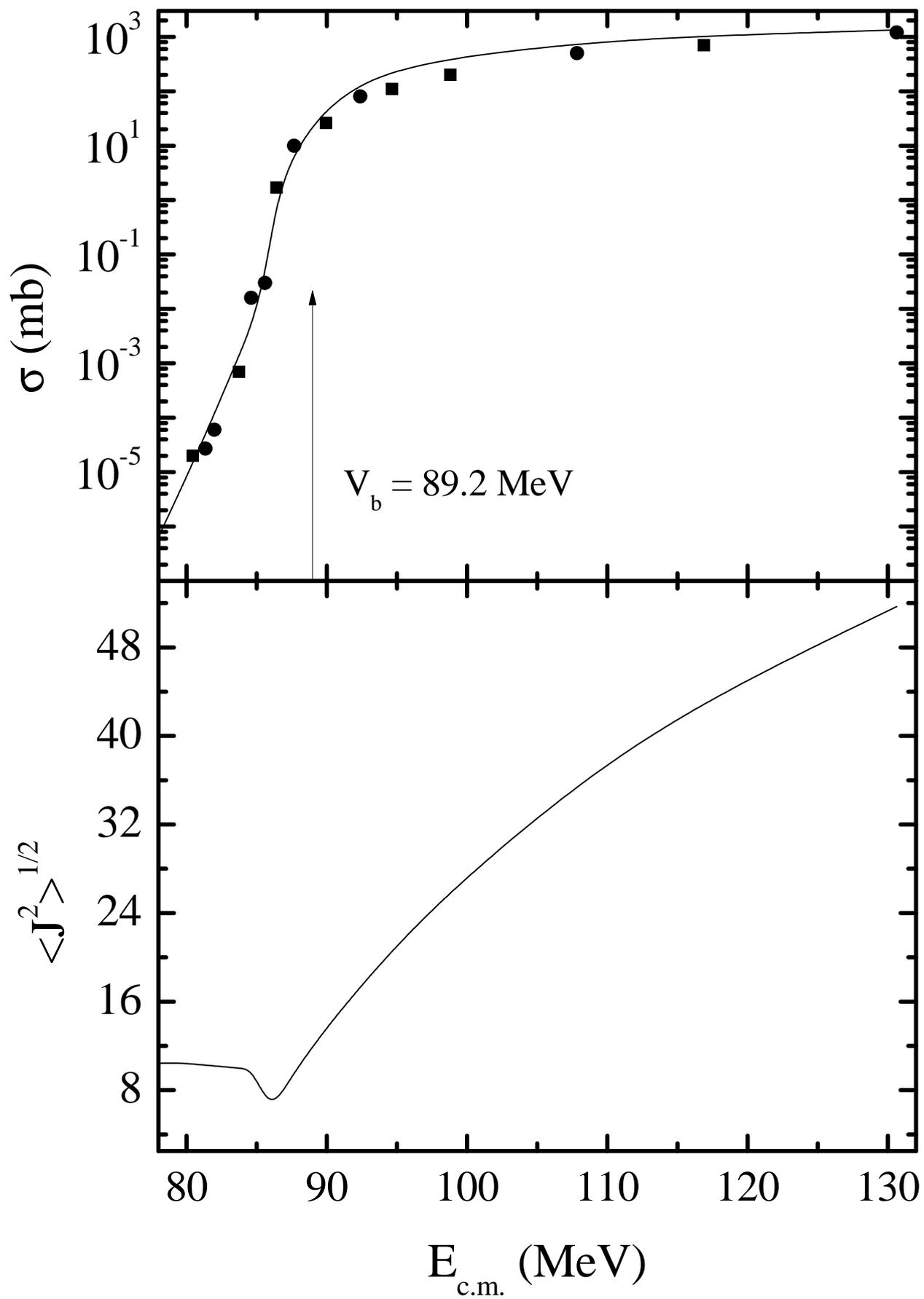
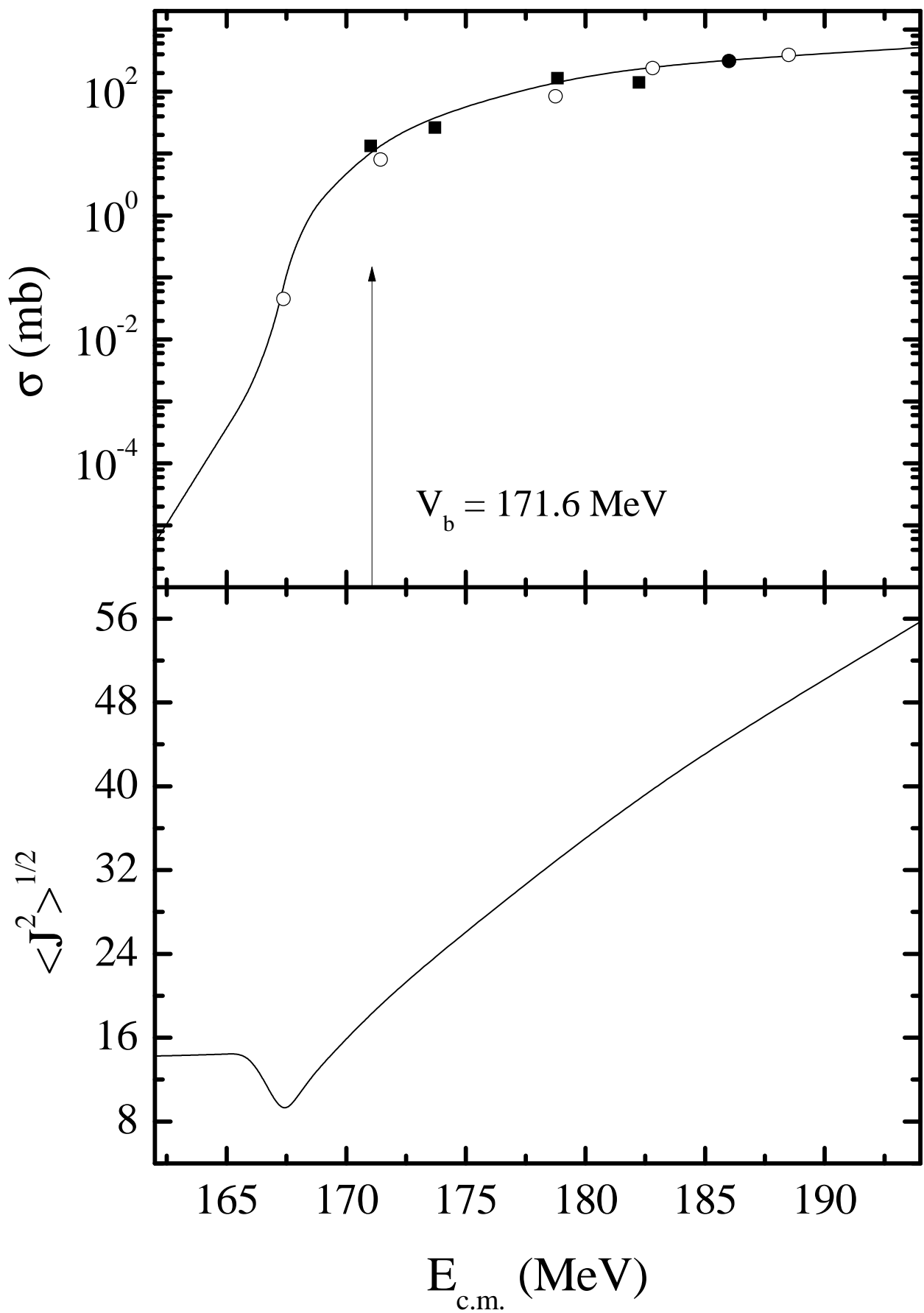
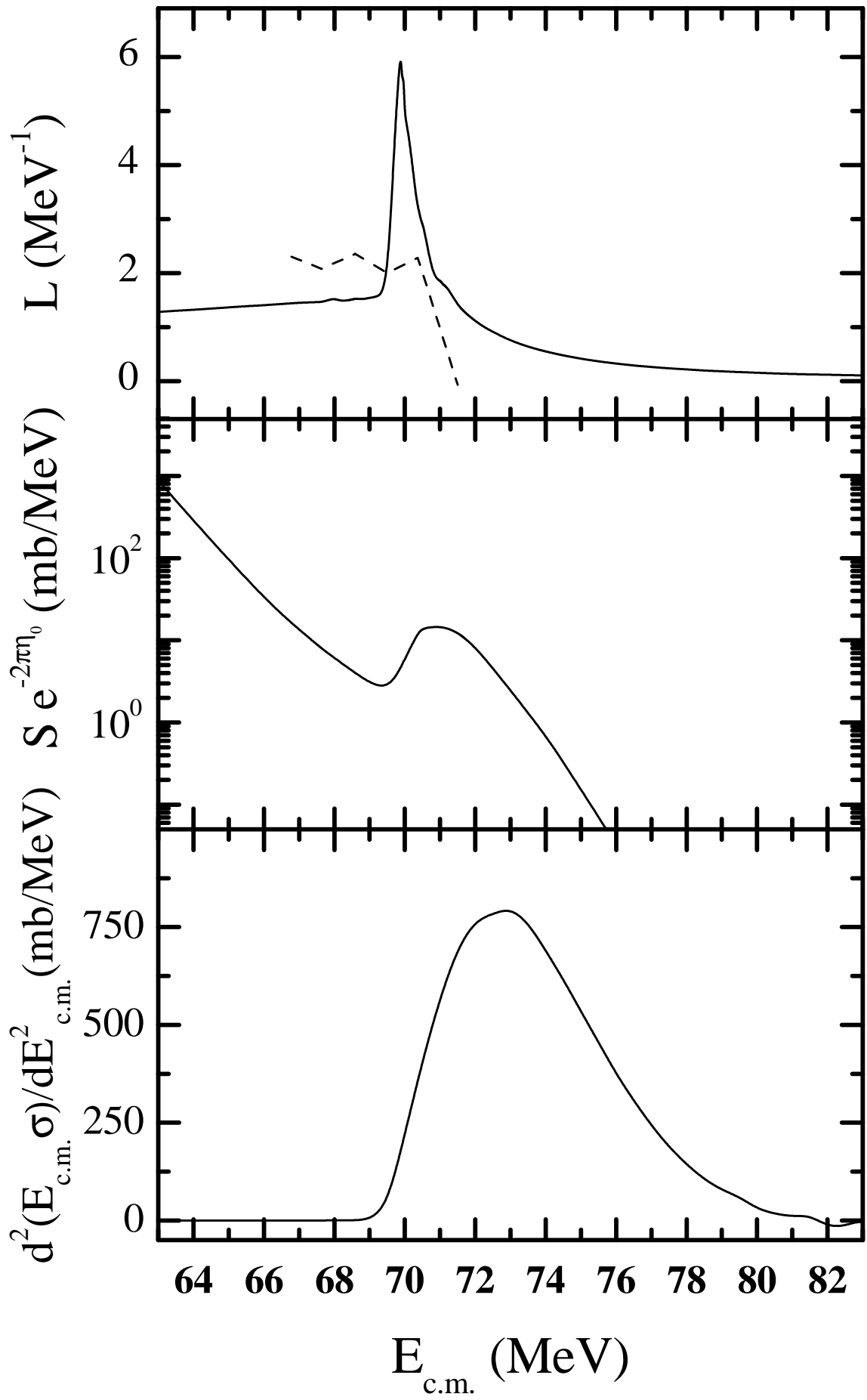
\end{document}